\documentclass[apl,twocolumn,bibnotes]{revtex4}

\usepackage{color,rotating,graphicx}
\usepackage{amsmath}
\usepackage{dsfont}
\usepackage{mathtools}
\renewcommand{\Im}{{\rm Im}}
\newcommand{\ri}{{\rm i}}
\newcommand{\re}{{\rm e}}
\newcommand{\rd}{{\rm d}}

\newcommand{\A}{{\rm A}}
\newcommand{\B}{{\rm B}}
\newcommand{\C}{{\rm C}}
\newcommand{\D}{{\rm D}}
\newcommand{\E}{{\rm E}}
\newcommand{\F}{{\rm F}}
\newcommand{\I}{{\rm i}}
\newcommand{\blockt}{\boldsymbol{T}^{-1}}

\begin{document}
\title{Topological near-field heat flow in a honeycomb lattice}

\date{\today}

\author{Annika Ott}
\author{Svend-Age Biehs}
\email{ s.age.biehs@uni-oldenburg.de} %% email address is required
\affiliation{Institut f\"{u}r Physik, Carl von Ossietzky Universit\"{a}t, D-26111 Oldenburg, Germany}

\begin{abstract}  
We study the near-field thermal radiation of topologically protected edge modes in a honeycomb lattice of plasmonic InSb nanoparticles. We show that the heat transport by near-field interaction is in the topological non-trivial phase dominated by the heat flux channel provided by the edge modes rather than the bulk modes. This heat flux channel allows for an enhanced heat transport along the edges of the honeycomb lattice. In particular for materials with relatively small dissipation we find a 30 to 50 times larger heat flux along the lattice edges than in the bulk.
\end{abstract}

\maketitle

%%%%%%%%%%%%%%%%%%%%%%%%%%%%%%%%%%%%%%%%%%%%%%%%%%%%%%%%%%%%%%%%%%%%%%%%%%%%%%%%%%%%%%%%%%%%%%%%%%%%%%%%%%%%%%%%%
%
% Introduction
%
%%%%%%%%%%%%%%%%%%%%%%%%%%%%%%%%%%%%%%%%%%%%%%%%%%%%%%%%%%%%%%%%%%%%%%%%%%%%%%%%%%%%%%%%%%%%%%%%%%%%%%%%%%%%%%%%%%%%%%

\section{Introduction} 

In the last decade, several theoretical interesting and even astonishing effects in the near-field radiative heat
transfer in many-body systems have been highlighted which have the potential for applications in nanotechnology.
To mention a few, it could be shown that non-reciprocal nanoparticle (NP) systems show a persistent heat current~\cite{silveirinha,zhufan,meinpaper,khandekar}, normal and anormal thermal Hall effects~\cite{hall,meinweylpaper}, giant magneto-resistance~\cite{giantlatella,bdrehen}. Furthermore, when the NPs interact with the surface modes of a nearby interface, then there can be an enhanced long-range heat transfer~\cite{SaaskilahtiEtAl2014,AsheichykEtAl2017,AsheichykEtAl2018,dong,paper_2sic}, a strong diode effect~\cite{aplpaper,meinspinpaper} when the interface supports non-reciprocal surface waves, and a strong directionality of heat transport and thermal emission due to hyperbolic or non-reciprocal plasmonic modes~\cite{zhang,filmpaper,he,YangEtAl2021,DongEtAl2021}. Reviews of these effects can be found in Refs.~\cite{mein_application,qnht,SongPlagiat2021,LatellaEtAl2021}. 

Recently, also plasmonic nanoparticle systems with a topological phase transition like a 1D Su-Schrieffer-Heeger chain~\cite{ungerade,acs,becssh} and a 2D Su-Schrieffer-Heeger lattice~\cite{sshgitter} where shown to have an important heat transfer and an enhanced near-field energy density due to the edge modes which exist in the topological non-trivial phase~\cite{energiedichtepaper}. Such topological insulator structures~\cite{TopologInsul} have greatly been studied in the context of topological photonics~\cite{TopologPhoton} including several transport processes~\cite{Transport2Dtop}. This also includes studies of heat transport as, for instance, the bosonic heat transport in a Hofstadter lattice~\cite{topologicalheatcurrents}. These studies are typically carried out in a open quantum system approach and refer to configurations which can typically be realized in well designed optical systems which meet the corresponding set of parameters needed to see the topological effects, like for example by using ultracold gases in optical lattices~\cite{topologicalheatcurrents}. 

In contrast to such studies based on open quantum systems we employ the theory of fluctuational electrodynamics which allows to describe the radiative heat flow in any configuration of dissipative materials at any given temperature or temperature distribution. In particular, we investigate the room temperature topological radiative heat flux in a honeycomb lattice (HL) of plasmonic NPs~\cite{arxiv,honey_top}. 

\begin{figure}[hbt!]
	\centering
	\includegraphics[width=0.48\textwidth]{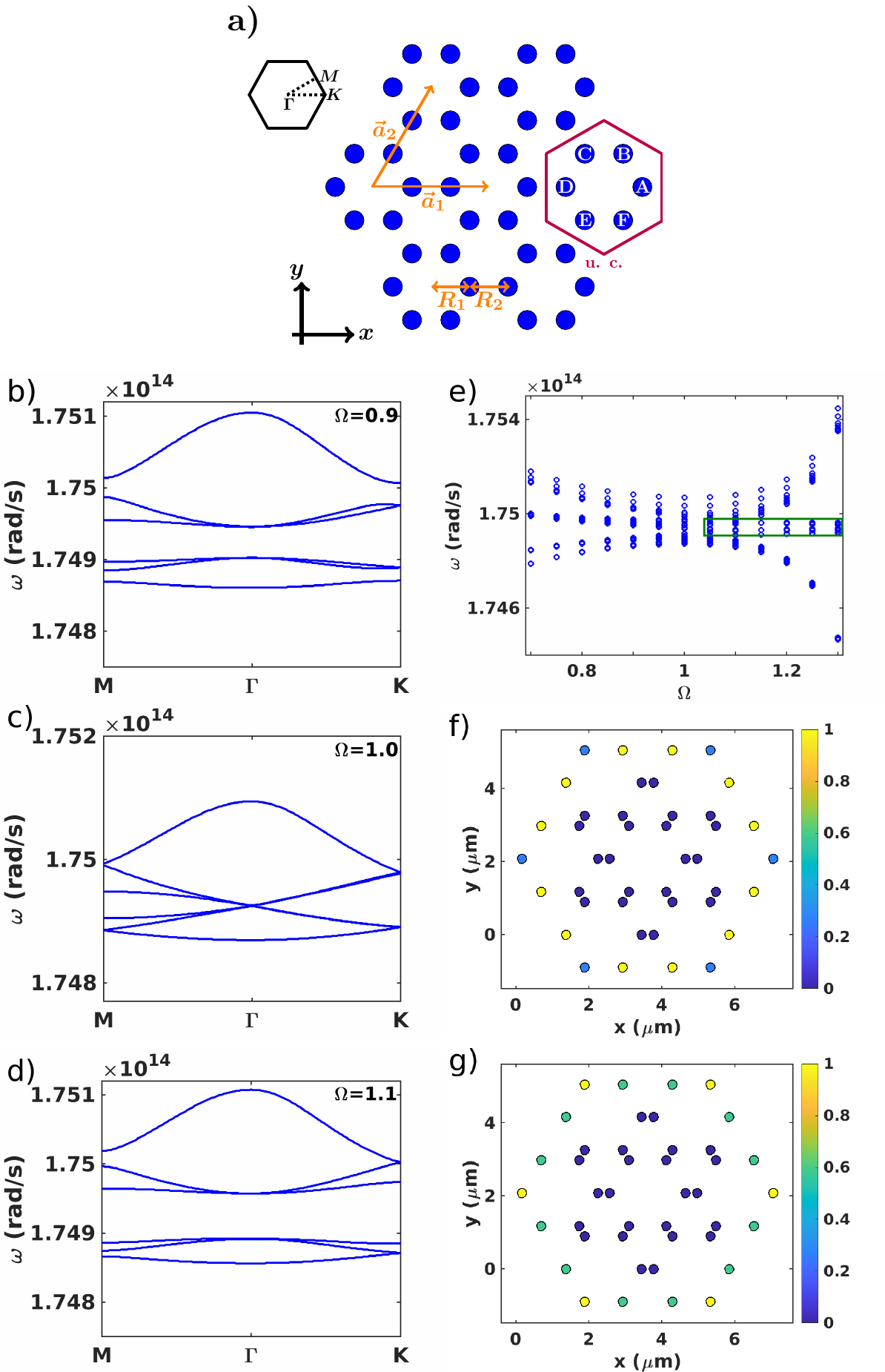} \\
	\caption{a) Sketch of the considered HL of InSb NPs in the x-y plane with lattice vectors $\vec{a}_1 = (d,0)^T$ and $\vec{a}_2 = \frac{d}{2}(1,\sqrt{3})^T$ and lattice constant $d$. As an example one unit cell (u. c.) is marked in red and the brillouin zone with its high symmetry points is shown. b)-d) Bandstructure for OP modes for different values of $\Omega$ in the quasi-static regime ($k_0d\ll1$). Only interactions between NPs with those of the next two other unit cells are taken into account. e) Eigen mode frequencies for OP modes of a HL of 42 NPs as function of $\Omega$. The topological edge and corner modes are enclosed with a green box. f,g) Absolute values of the normalized dipole moments of the OP modes at the NP's positions for the edge mode frequency ${\omega_{\rm op}=1.74931\cdot10^{14}}$~rad/s (top) and for the corner mode frequency $\omega_{\rm op}=1.74921\cdot10^{14}$~rad/s (bottom).}
	\label{konfiguration}
\end{figure}

%%%%%%%%%%%%%%%%%%%%%%%%%%%%%%%%%%%%%%%%%%%%%%%%%%%%%%%%%%%%%%%%%%%%%%%%%%%%%%%%%%%
%
%  Nanoparticles
%
%%%%%%%%%%%%%%%%%%%%%%%%%%%%%%%%%%%%%%%%%%%%%%%%%%%%%%%%%%%%%%%%%%%%%%%%%%%%%%%%%%%
\section{Theoretical Model}

In the following, we chose a HL configuration as sketched in Fig.~\ref{konfiguration}a). Each unit cell consists of six NPs belonging to the sublattices $\A$-$\F$. The lattice vectors $\vec{a}_1$ and $\vec{a}_2$ determine the periodicity of the HL with the lattice constant $d$ in both directions given by their length. We introduce the lattice parameter $\Omega \in [0,\frac{3}{2}]$ to determine the distances $R_1$ and $R_2$ in the lattice which are given in terms of the lattice constant $d$ by $R_1 = \Omega\frac{d}{3}$ and $R_2=d-2R_1$. If $\Omega=1$ we have $R_1 = R_2 = \frac{d}{3}$ the HL is ``normal'', if $\Omega<1$ ($R_1 < R_2$) it is shrunken and if $\Omega>1$ ($R_1 > R_2$) it is expanded. The HL has $C_6$ and inversion symmetry~\cite{symmhoneyurquelle}.

The NP in the lattice are spherical InSb NPs with a radius of $R=100$~nm and a lattice constant of $d=24R$. Since the peak wavelength of thermal radiation is at room temperature about $10\mu$m and $d \gg R$ we can use the dipole approximation to describe the electromagnetic response of the NPs~\cite{dipolapprox}. The polarizability of all the NPs in the lattice is then given by~\cite{qalpha1,qalpha2}
\begin{equation} 
  \alpha = 4\pi R^3 \frac{\epsilon - 1}{\epsilon + 2}
\end{equation} 
with the permittivity of highly doped InSb described by a Drude model~\cite{qeps}
\begin{equation}
 	\epsilon = \epsilon_\infty \left(1 - \frac{\omega_{\rm p}^2}{\omega(\omega+{\rm i}\gamma)} \right)
\end{equation}
using the damping constant $\gamma=1\cdot10^{12}$~rad/s, the high-frequency dielectric constant $\epsilon_\infty=15.68$, the density of the free charge carriers $n=1.36\cdot10^{19}$~cm$^{-3}$, and the effective mass $m^*=7.29\cdot10^{-32}$~kg so that the plasma frequency is $\omega_{\rm p} = 1.86\times10^{14}\,{\rm rad/s}$ and the localized NP resonance is at $\omega_{\rm LPR} = 1.752\times10^{14}\,{\rm rad/s}$. 

%%%%%%%%%%%%%%%%%%%%%%%%%%%%%%%%%%%%%%%%%%%%%%%%%%%%%%%%%%%%%%%%%%%%%%%%%%%%%%%%%%%
%
%  Bloch ansatz
%
%%%%%%%%%%%%%%%%%%%%%%%%%%%%%%%%%%%%%%%%%%%%%%%%%%%%%%%%%%%%%%%%%%%%%%%%%%%%%%%%%%%

The eigenmodes of the HL can be determined by considering for example the induced dipole moment $\vec{p}_{\delta_m} = \epsilon_0 \alpha \vec{E}(\mathbf{r}_{\delta_m})$ of the NP $\delta \in\{\A, \B, \C, \D, \E, \F\} $ in the m-th unit cell at position $\mathbf{r}_{\delta_m}$ due to the field $\vec{E}$ of all the other dipoles. For an infinite lattice the unit cells are all equivalent and a Bloch ansatz for the dipole moments can be made to derive a $6\times3$-dimensional eigenvalue equation given by
\begin{equation}
  \Bigg(\boldsymbol{M}-\frac{1}{\alpha}\boldsymbol{1}\Bigg)\boldsymbol{\vec{p}}=\boldsymbol{\vec{0}}
\label{disph}
\end{equation}
where $\boldsymbol{\vec{p}}=(\vec{p}_{\A}, \vec{p}_{\B},\vec{p}_{\C}, \vec{p}_{\D}, \vec{p}_{\E}, \vec{p}_{\F})^T$ is the periodic part of the Bloch expansion. The $\gamma\delta$ component with $\gamma, \delta\in\{\A, \B, \C, \D, \E, \F\}$ of the ``Hamiltonian'' for an arbitrary unit cell $m$ is given by
\begin{equation}
   \boldsymbol{M}_{\gamma\delta} = k_0^2\sum_{j}(1-\delta_{ \gamma_m\delta_j})\mathds{G}^{{\rm E}}_{\gamma_{m}\delta_{j}}e^{\I \vec{k} \cdot (\vec{r}_{\delta_j}-\vec{r}_{\gamma_{m}})}
\label{mkomph}
\end{equation}
where $k_0=\omega/c$ is the vacuum wave vector, $c$ is the light velocity in vacuum, and $\mathds{G}_{\gamma_m \delta_j}^{\rm E} \equiv \mathds{G}^{\rm E}(\vec{r}_{\gamma_m},\vec{r}_{\delta_j})$ is the electric Green function~\cite{novotny} {which can be written as
\begin{equation}
   \mathds{G}^{\rm E}(\vec{r}_{\gamma_m},\vec{r}_{\delta_j}) = \frac{\re^{\ri k_0 d_{\gamma \delta}}}{4 \pi d_{\gamma \delta}} \bigl[ a \mathds{1} + b \vec{e}_{\gamma\delta} \otimes \vec{e}_{\gamma\delta} \bigr].
\end{equation}
Here $d_{\gamma \delta} = |\vec{r}_{\gamma_m} - \vec{r}_{\delta_j}|$ is the distance between particle $\gamma_m$ and $\delta_j$ and $\vec{e}_{\gamma\delta} = (\vec{r}_{\gamma_m} - \vec{r}_{\delta_j})/d_{\gamma \delta}$ is the unit vector connecting both particles. The symbol $\mathds{1}$ stands for the $3\times3$ unit matrix and $\otimes$ stands for the tensor product. The functions $a$ and $b$ are defined as
\begin{equation}
  a = 1 + \frac{\ri k_0 d_{\gamma\delta} - 1}{k_0^2 d_{\gamma\delta}^2} \quad \text{and} \quad b = \frac{3 - 3\ri k_0 d_{\gamma\delta} - k_0^2 d_{\gamma\delta}^2}{k_0^2 d_{\gamma\delta}^2}.
\end{equation}
From this representation it becomes clear that when the particles are within the x-y plane then the Green function has the symmetry
\begin{equation}
  \mathds{G}^{\rm E} = \begin{pmatrix} G^{\rm E}_{xx} & G^{\rm E}_{xy} & 0 \\ G^{\rm E}_{yx} & G^{\rm E}_{yy} & 0 \\ 0 & 0 &  G^{\rm E}_{zz} \end{pmatrix} 
\end{equation}
where we suppressed the arguments for convenience. This symmetry suggests that the in-plane (IP) and out-of-plane (OP) modes are decoupled so that their band structures can be evaluated separately. Therefore, when calculating properties of the IP modes only we restrict the Green function to the two-dimensional $x$-$y$ subspace such that we consider in this case
\begin{equation}
\begin{split}
   \mathds{G}^{\rm E, IP} &= \begin{pmatrix} G^{\rm E}_{xx} & G^{\rm E}_{xy} \\ G^{\rm E}_{yx} & G^{\rm E}_{yy} \end{pmatrix} \\
                          &= \frac{\re^{\ri k_0 d_{\gamma \delta}}}{4 \pi d_{\gamma \delta}} \bigl[ a \mathds{1}_{2\times2} + b \vec{e}_{\gamma\delta} \otimes \vec{e}_{\gamma\delta} \bigr]
\end{split}
\end{equation} 
with the $2\times2$ unit matrix $\mathds{1}_{2\times2}$. Consequently, each block matrix element $\boldsymbol{M}_{\gamma\delta}^{\rm IP}$ in Eq.~(\ref{mkomph}) is therefore given by a $2\times2$ matrix. Likewise, when calculating properties for the OP modes we restrict the Green function to the one-dimensional $z$ subspace considering
\begin{equation}
   G^{\rm E, OP} = G^{\rm E}_{zz} = \frac{\re^{\ri k_0 d_{\gamma \delta}}}{4 \pi d_{\gamma \delta}} a.
\end{equation} 
In this case each block matrix element $\boldsymbol{M}_{\gamma\delta}^{\rm OP}$ is a scalar. The corresponding eigenvalue equations (\ref{disph}) are then $6\times2$ dimensional for IP and $6\times1$ dimensional for OP modes. Hence, in general for the IP modes 12 bands can form whereas for the OP modes only 6 bands can be expected.  
}

It can be shown that as in the Su-Schrieffer-Heeger model the Hamiltonian $\mathbf{M}$ in the above model has a chiral symmetry if we consider interaction between the nearest neighbours in the quasi-static regime ($k_0d\ll1$) only, which is together with the inversion symmetry responsible for the appearance of topologically protected edge states in finite lattices \cite{topinvm}. In Fig.~\ref{konfiguration}b)-d) the bandstructures for the OP modes for different values of $\Omega$ are shown. The results for the IP modes are similar. It can be seen that the bands in the middle are closed at the $\Gamma$ point for $\Omega=1$ showing the typical Dirac-like linear dispersion~\cite{prl}. For $\Omega\neq1$ a band gap evolves. In these band gaps we will find topologically protected edge and corner modes for certain values of $\Omega$.    

%%%%%%%%%%%%%%%%%%%%%%%%%%%%%%%%%%%%%%%%%%%%%%%%%%%%%%%%%%%%%%%%%%%%%%%%%%%%%%%%%%%
%
%  Zak phase
%
%%%%%%%%%%%%%%%%%%%%%%%%%%%%%%%%%%%%%%%%%%%%%%%%%%%%%%%%%%%%%%%%%%%%%%%%%%%%%%%%%%%
To know for which values these edge modes appear, we will determine the Zak phase of the 2D lattice. In the quasi-static regime ($k_0d\ll1$) a 2D Zak phase for which we introduce the notation $\vec{Z} = \sum_{i=1}^{2}Z_i\vec{a}_i\eqqcolon (Z_1,Z_2)^T$ can be defined~\cite{2dzakoriginal,paritaet_allg}. Due to inversion symmetry only the time reversal invariant momenta (TRIM) are relevant for determining the Zak phase~\cite{topinvm}. Here, the TRIM are of course only real TRIM if retardation is neglected as it is the case in the quasi-static regime. Thus, for the HL the $i$-th component of the Zak phase can be written as~\cite{paritaet_allg,liutop} 
\begin{equation}
  Z_i=\Bigg(\sum_{n=1}^{\tilde{N}} Z_i^n\Bigg) \mod 1 \label{glph}
\end{equation}
where $\tilde{N}$ is the number of bands below the band gap and the Zak phase component in band $n$ is
$Z_i^n=\pi q_i^n$ modulo $2 \pi$ with $q_i^n$ defined via the symmetry properties at the TRIM by
\begin{equation}
  (-1)^{q_i^n}=\frac{\eta^n(M_i)}{\eta^n(\Gamma)}.
\end{equation}
Here, $\eta$ represents the parity under $C_2$ rotation at the TRIM $M_1=\pi/d(1,-1/\sqrt{3})^T$ and $M_2=\pi/d(0,2/\sqrt{3})^T$. Note that due to the $C_6$ symmetry we have $Z_1^n=Z_2^n$~\cite{paritaethoney}. A second topological invariant which is called the quadrupole polarization $Q_{12}$ in analogy to the electronic band theory~\cite{dipolquad} can be defined via~\cite{paritaethoney,verstemarxiv,quelleb}
\begin{eqnarray}
  Q_{12}=\frac{1}{4\pi^2}\Bigg(\sum_{n=1}^{\tilde{N}}Z_1^nZ_2^n\Bigg) \mod 1\label{glqh}.
\end{eqnarray}
With Eq.~(\ref{glph}) and Eq.~(\ref{glqh}) for $\tilde{N}=3$ the Zak phase in the band gap in Fig.~\ref{konfiguration}b)-c) is $\vec{Z}=(0,0)^T$ and the quadrupole polarisation is $Q_{12}=0$ for $\Omega<1$ for OP and IP modes. On the other hand, for $\Omega>1$ we obtain $\vec{Z}=(0,0)^T$ and $Q_{12}=1/2$ in agreement with Refs.~\cite{paritaethoney,liutop}. The result indicates that the lattice is in a topological trivial phase if $\Omega<1$ and in the topological non-trivial phase if $\Omega>1$. Hence, the bulk-edge correspondence predicts topological edge and corner modes in a finite HL for $\Omega>1$~\cite{paritaethoney}.

%%%%%%%%%%%%%%%%%%%%%%%%%%%%%%%%%%%%%%%%%%%%%%%%%%%%%%%%%%%%%%%%%%%%%%%%%%%%%%%%%%%
%
%  Finite Lattice
%
%%%%%%%%%%%%%%%%%%%%%%%%%%%%%%%%%%%%%%%%%%%%%%%%%%%%%%%%%%%%%%%%%%%%%%%%%%%%%%%%%%%

For a finite HL of $N$ NPs we consider first the induced dipole moment 
\begin{equation}
  \vec{p}^{\rm ind}_i = \epsilon_0 \alpha \vec{E}(\vec{r}_i)
\label{Eq:pind}
\end{equation}
at position $\vec{r}_i$ due to the electric field $\vec{E}(\vec{r}_i)$ due to all other $N - 1$ dipoles; $\epsilon_0$ is the vacuum permittivity. Since, the field of all other dipoles is determined via the electric Green function by
\begin{equation}
  \vec{E}(\vec{r}_i) = k_0^2 \alpha \sum_{j \neq i} \mathds{G}^{\rm E}(\mathbf{r}_i, \mathbf{r}_j) \mathbf{p}_j
\end{equation}
we obtain by inserting this expression in Eq.~(\ref{Eq:pind}) the general $N$ NP eigenvalue equation~\cite{nteilchen}
\begin{equation}
	\boldsymbol{T}\cdot\boldsymbol{p}=\boldsymbol{{0}}
	\label{block}
\end{equation}
with the $3N$ dimensional block vector $\boldsymbol{p}$ containing the $N$ dipole moments $\vec{p}_i$ of each NP $i$
and the $3N \times 3N$ block matrix with the components 
\begin{equation}
  \boldsymbol{T}_{ij} = \delta_{ij}\mathds{1}-(1-\delta_{ij})k_0^2\alpha\mathds{G}_{ij}^{\rm E}.
  \label{Eq:Tij}
\end{equation}
By positioning the $N$ NPs in a HL we can determine the eigenmode frequencies for the corresponding finite structure by evaluating $\det(\boldsymbol{T}) = 0$. On the other hand, a Bloch ansatz for the dipole moments in Eq.~(\ref{block}) leads to Eq.~(\ref{disph}).

In Fig.~\ref{konfiguration}e) the eigen mode frequencies for the OP modes for a lattice of 42 InSb NPs are shown for different values of $\Omega$. For $\Omega>1$ it can be seen that topological edge and corner modes appear in the band gap. To clearly see that indeed we have here edge and corner modes, we show the absolute value of the eigen dipole moments of each NP for the edge and corner modes frequencies in Fig.~\ref{konfiguration} f)-g) for $\Omega=1.3$ and the OP modes.  The behaviour for IP modes is not shown here, because it is qualitatively the same. There, the absolute values of the dipole moments of the edge (corner) NPs dominate if an edge (corner) mode frequency is used. Due to dissipation and a related overlap of the edge and corner mode frequencies due to their close spectral location the edge (corner) dipole moments are also relatively large when a corner (edge) mode is excited. As expected, the edge (corner) modes are well localized at the edges (coners).

\section{Results and Discussion}

\begin{figure}[hbt!]
	\centering
	\includegraphics[width=0.35\textwidth]{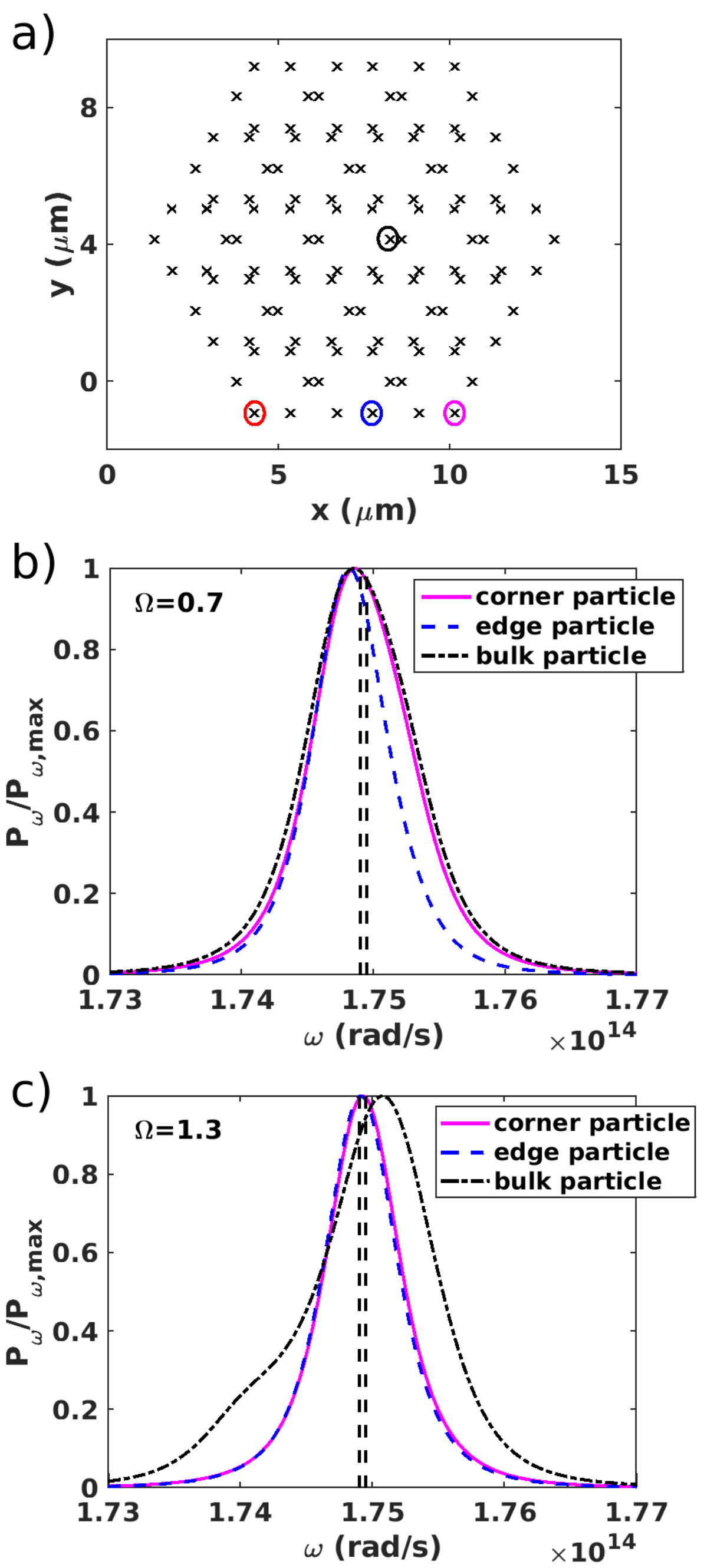} \\
	\caption{Spectral net heat transfer: a) configuration: Black points show the NP positions of a lattice with 114 NPs and $\Omega=1.3$. The NP at the red marked position has a temperature of 350 K whereas the others and the background have 300 K. The net heat transfer on the NP of the blue marked position (edge NP), on the NP at the pink marked position (corner NP) and on the NP at the black marked position (bulk NP) is investigated. b) Spectral net heat transfer for $\Omega = 0.7$ and c) $\Omega = 1.3$. The dominating topological frequencies $\omega= 1.7495\cdot10^{14}$ rad/s for OP and ${\omega= 1.749\cdot10^{14}}$~rad/s for IP modes is indicated by a dashed line.}
	\label{nht_spek}
\end{figure}

We study the transfered power between different NPs in a HL of $N = 114$ NPs as indicated in Fig.~\ref{nht_spek}. The net transferred power from the hot edge NP (marked by a red circle) in  Fig.~\ref{nht_spek} to the marked edge, corner and bulk NPs labeled as NPs $i$ is determined by using the expression~\cite{nteilchen} 
\begin{equation}
  P_i =  \int_0^\infty \!\! \frac{\rd \omega}{2 \pi}\, P_{i,\omega} = 3 \int_0^\infty \!\! \frac{\rd \omega}{2 \pi}\sum_{j=1}^N\bigl[\Theta(T_j) - \Theta(T_i)\bigr] \mathcal{T}_{ji} \label{nht}
\end{equation}
with the Bose-Einstein function $\Theta(T) = \hbar \omega / (\exp(\hbar \omega/ k_B T) - 1)$, the reduced Planck constant $\hbar$
 and the Boltzmann constant $k_B$. Furthermore, the transmission coefficient $\mathcal{T}_{ji}$ for the power exchanged between the NPs is given by~\cite{nteilchen}
\begin{align}
\mathcal{T}_{ji} &= \frac{4}{3}\frac{{\rm Im}(\alpha)^2}{|\alpha|^2}\blockt_{ij}{\blockt_{ij}}^\dagger
\end{align}
using the same block matrix $\mathbf{T}$ as in Eq.~(\ref{block}). In Eq.~({\ref{nht}}) we neglected the heat transfer to the background. This is possible because we consider a stationary case with NP distances much smaller then the thermal wavelength, so that this contribution is negligible small compared to the interparticle heat flux~\cite{nteilchen}. In Fig.~\ref{nht_spek} the spectra for the net heat transfer are shown. It can be seen that in the topological trivial case all maxima are located in the bands of the bulk modes below the bandgap. In contrast, for the topological non-trivial case the maximum for the net heat transfer to the edge (corner) NPs is at the corner (edge) modes frequencies whereas for the bulk NP it is at frequencies in the bands of the bulk modes above the band gap. Due to dissipation edge and corner modes frequencies overlap and thus cannot be distinguished without strongly reducing the damping constant. Thus, heating a corner NP instead of an edge NP will result in the same effect that the other edge/corner NPs are heated due to the heat flux channel provided by the edge modes, whereas the bulk NPs are heated via the channel provided by the bulk modes. 

\begin{figure}[hbt!]
	\centering
	\includegraphics[width=0.45\textwidth]{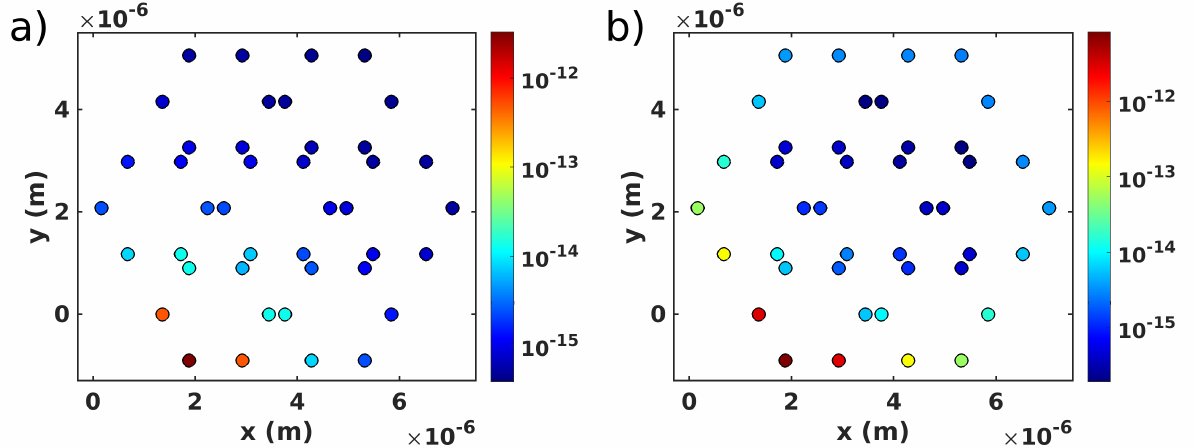} \\
	\caption{Integrated heat transfer: a) absolute value of net transferred power $P_i$ at the position of each NP for the "natural" $\gamma=10^{12}$~rad/s b) and the reduced damping constant $\gamma=10^{11}$~rad/s (bottom). The particle in the left bottom corner has a temperature of 350 K whereas the other and the background have 300 K.  %(right) Relevant part of the spectral poynting vector $\langle \vec{S}^{\rm tr}_{\omega}\rangle$ (W/m$^2$) at edge mode frequency $\omega= 1.7495\cdot10^{14}$~rad/s for $\gamma=10^{12}$~rad/s (top) and for $\gamma=10^{11}$~rad/s (bottom). Furthermore the absolute value of the in-plany Poynting vector components are shown by the colorbar.
	}
	\label{nht_ort}
\end{figure}

%%%%%%%%%%%%%%%%%%%%%%%%%%%%%%%%%%%%%%%%%%%%%%%%%%%%%%%%%%%%%%%%%%%%%%%%%%%%%%%%%%%
%
%  Poynting Vector
%
%%%%%%%%%%%%%%%%%%%%%%%%%%%%%%%%%%%%%%%%%%%%%%%%%%%%%%%%%%%%%%%%%%%%%%%%%%%%%%%%%%%
To have a better understanding of the pathways of the heat flux in the topological non-trivial case, we show in Fig.~\ref{nht_ort} the fully integrated transferred power $P_i$ at the location of each NP when the NP at the left bottom corner is heated to $350\,{\rm K}$ as in the configuration in Fig.~\ref{nht_spek} but for a slightly smaller lattice with $N = 42$ NPs. When choosing the  "natural" damping constant $\gamma=10^{12}$~rad/s for InSb then it can be hardly seen that the heat is dominantly flowing along the edges. On the other hand, if we reduce the damping constant by a factor of 10 the preferred heat transfer on the edge and corner NPs can be nicely seen. In Fig.~\ref{nht_ort} we also show the relevant part of the mean Poynting vector  ($\alpha = x,y,z$)
  \begin{equation}
  \begin{split}
        \langle S_{\omega,\alpha}^{\rm tr} (\vec{r}) \rangle &= 4\omega\mu_0k_0^2 \sum_{\beta,\gamma = x,y,z}\epsilon_{\alpha\beta\gamma}\sum_{ijk=1}^{N}\bigl[ \Theta(T_j) - \Theta(T_b) \bigr] \\
                                                   & \times \Im(\alpha) {\rm Re}\Big[ \mathds{G}^{\rm E}(\vec{r},\vec{r}_i)\blockt_{ij} (\mathds{G}^{\rm H} (\vec{r},\vec{r}_k)\blockt_{kj})^\dagger\Big]_{\beta\gamma}.
  \end{split}
  \label{poyntingsurf}
\end{equation} 
describing the heat flux between the NPs as discussed in Ref.~\cite{meinspinpaper}. Here, $\mu_0$ is the permeability of vacuum and $T_b$ is the temperature of the background which is assumed to be $300\,{\rm K}$, and $\mathds{G}^{\rm H} = \frac{1}{\ri \omega \mu_0}\nabla\times\mathds{G}^{\rm E}$ is the Green function of the magnetic field due to the electric dipoles. Again, it can be seen that for the "natural" dissipation the larger heat flux at the edge and corner NPs cannot be clearly seen but for the reduced damping by a factor of 10 the heat flux along the edge and corners is clearly dominating.

\begin{figure}[hbt!]
	\centering
	\includegraphics[width=0.4\textwidth]{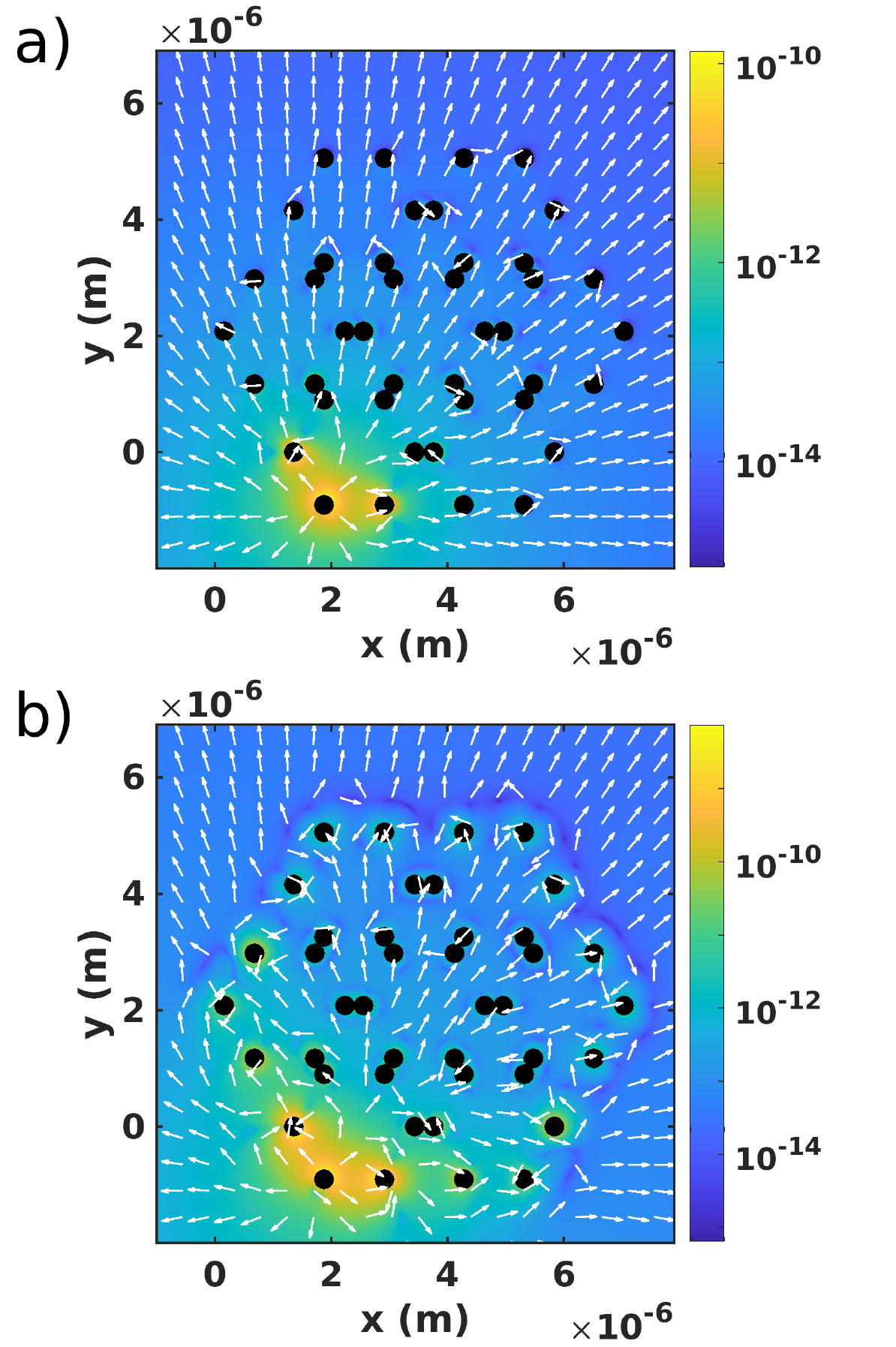} \\
	\caption{%Integrated heat transfer: (left) net transferred power $P_i$ at the position of each NP for the "natural" $\gamma=10^{12}$~rad/s (top) and the reduced damping constant $\gamma=10^{11}$~rad/s (bottom). (right) 
	Relevant part of the spectral Poynting vector $\langle \vec{S}^{\rm tr}_{\omega}\rangle$ (Ws/m$^2$) at edge mode frequency $\omega= 1.7495\cdot10^{14}$~rad/s for a) $\gamma=10^{12}$~rad/s and b) for $\gamma=10^{11}$~rad/s. Furthermore the absolute value of the in-plane Poynting vector components are shown by the colorbar. The temperatures are as in Fig.~\ref{nht_ort}.}
	\label{poynting_ort}
\end{figure}

%%%%%%%%%%%%%%%%%%%%%%%%%%%%%%%%%%%%%%%%%%%%%%%%%%%%%%%%%%%%%%%%%%%%%%%%%%%%%%%%%%%
%
% Damping 
%
%%%%%%%%%%%%%%%%%%%%%%%%%%%%%%%%%%%%%%%%%%%%%%%%%%%%%%%%%%%%%%%%%%%%%%%%%%%%%%%%%%%

\begin{figure}[hbt!]
	\centering
	\includegraphics[width=0.4\textwidth]{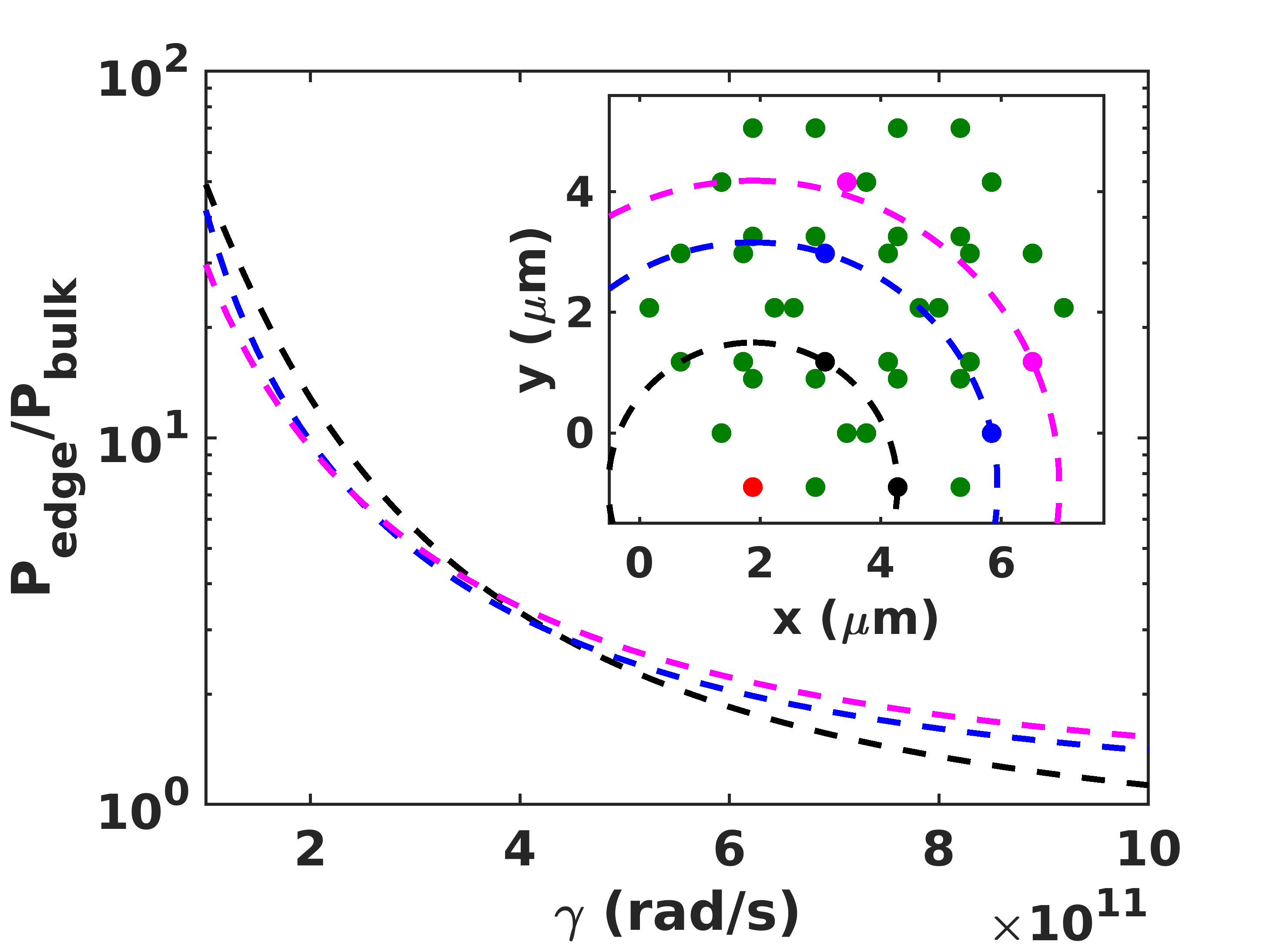}
	\caption{Ratio $P_{\rm edge}/P_{\rm bulk}$ of the fully transmitted power to the edge and bulk NP as function of the damping constant of InSb. (Inset) Honeycomb lattice with $N = 42$ and $\Omega=1.3$. The red NP has a temperature of 350 K while the others and the background have a temperature of 300 K. The NPs on the black, blue, and magenta circles have the same distance to the hot edge NP. To calculate $P_{\rm edge}$ ($P_{\rm bulk}$) the transferred power to the black, blue, and magenta edge (bulk) NP is calculated. }
	\label{nht_pepb}
\end{figure}

From the above results it becomes clear that the damping constant is the limiting factor for a topological edge mode dominated heat flux. This impact is even more obvious in Fig.~\ref{nht_pepb}. There, we show the ratio $P_{\rm edge}/P_{\rm bulk}$ of the transferred power to an edge NP compared to that to a bulk NP with nearly the same distance to the hot corner NP as indicated in the inset of Fig.~\ref{nht_pepb}. By reducing the damping constant the spectral overlap between the edge and bulk modes is reduced so that the two channels are clearly separated. As a result $P_{\rm edge}/P_{\rm bulk}$ is increasing with decreasing values of the damping constant. This confirms the edge mode domination of the net heat transfer in the topological non-trivial case. But even with the ``natural'' damping constant of InSb $P_{\rm edge}/P_{\rm bulk}$ is 1.129 for the power transmitted to the NP in the black circle, 1.4 for those in the blue circle and 1.528 for the NP in the magenta circle. Hence, even for the ``natural'' damping constant the net heat transfer to the edge NPs is higher than to the bulk NPs. A strong dominance of the edge modes by a factor of 30 to 50 can be seen when choosing a reduced damping constant.

 We have also calculated $P_{\rm edge}$ and $P_{\rm bulk}$ for other materials as listed in table~\ref{Tab:Mat}. For SiO$_2$ we use the optical data from Ref.~\cite{Palik}, for GaN and SiC we employ the Drude-Lorentz model with parameters as in Refs.~\cite{nteilchen,Barker}, and for the uni-axial hBN we use the model from Ref.~\cite{filmpaper}. It can be seen that for the here tested materials $P_{\rm edge}/P_{\rm bulk}$ is typically between $1$ and $1.6$. For all materials $P_{\rm edge}/P_{\rm bulk}$ increases when decreasing the damping so that for low damping a larger $P_{\rm edge}/P_{\rm bulk}$ can be expected. For example, for hBN we have $\gamma_{\rm hBN} = 0.75\times10^{12}\,{\rm rad/s}$, for SiC we use $\gamma_{\rm SiC} = 0.9\times10^{12}\,{\rm rad/s}$, and for GaN we have $\gamma_{\rm GaN} = 1.59\times10^{12}\,{\rm rad/s}$. Consequently, we find the largest value of $P_{\rm edge}/P_{\rm bulk}$ for hBN and the smallest for GaN. More detailed studies for different materials and configurations are needed to get further insight in the possibilities to optimize the edge mode heat transfer.

\begin{table}
\begin{tabular}{|c|c|c|c|}
  \hline
  material   & $P_{\rm edge}$ ($10^{-16}$ W) & $P_{\rm bulk}$ ($10^{-16}$ W)  & $P_{\rm edge}/P_{\rm bulk}$  \\ \hline \hline
  InSb       & 7.92  & 5.18 & 1.53 \\ \hline
  SiO$_2$    & 0.79  & 0.71 & 1.1  \\ \hline
  GaN        & 1.51  & 1.27 & 1.19 \\ \hline
  SiC        & 2.9   & 2.28 & 1.27 \\ \hline
  hBN (OA,x) & 3.51  & 2.63 & 1.33 \\ \hline   
  hBN (OA,y) & 7.35  & 4.7 & 1.56 \\ \hline   
%  B4C	   & 33.2  & 11.3 & 2.9  \\ \hline
\end{tabular}
\caption{\label{Tab:Mat} Values for $P_{\rm edge}$, $P_{\rm bulk}$, and $P_{\rm edge}/P_{\rm bulk}$ for the transmitted power to the magenta edge or bulk particle in Fig.~\ref{nht_pepb} using different materials for the NPs in the HL. For hBN we show the results for the optical axis (OA) in $y$ or $x$ direction.}
\end{table}

%%%%%%%%%%%%%%%%%%%%%%%%%%%%%%%%%%%%%%%%%%%%%%%%%%%%%%%%%%%%%%%%%%%%%%%%%%%%%%%%%%%%%%%%%%%%%%%%%%%%%%%%%%%%%%%%%
%
% Conclusion
%
%%%%%%%%%%%%%%%%%%%%%%%%%%%%%%%%%%%%%%%%%%%%%%%%%%%%%%%%%%%%%%%%%%%%%%%%%%%%%%%%%%%%%%%%%%%%%%%%%%%%%%%%%%%%%%%%%%%%%%
\section{Conclusions}

In conclusion we have shown that the edge and corner modes in a 2D plasmonic HL provide the dominant radiative heat flux channel for near-field thermal interaction when the HL is in its non-trivial phase. This heat flux channel allows for a preferred heat flow along the edges of the lattice so that in this phase the bulk of the HL is in some sense thermally isolated from its edge. This effect can be used to control the pathways of the heat flux in more complex two-dimensional structures or meta-surfaces. We have furthermore shown that the separation between the bulk mode and edge mode dominated heat flux channels is highly dependent on the damping constant, i.e.\ the dissipative properties, of the used building blocks of the 2D structures which were in our case InSb NPs. Here, in particular relatively small damping favors the dominance of the edge mode heat flux channel so that for any application of these findings corresponding materials with low damping are preferable. Furthermore, by adding a magnetic field to the InSb HL or by using Weyl semi-metals one can expect to have circular persistent heat fluxes with angular momentum and spin texture in the HL as found in Refs.~\cite{silveirinha,zhufan,hall,meinpaper,khandekar,meinweylpaper} and therefore we expect to have a uni-directional heat transfer by the edge modes in this case. This possibility will be investigated elsewhere.   

%%%%%%%%%%%%%%%%%%%%%%%%%%%%%%%%%%%%%%%%%%%%%%%%%%%%%%%%%%%%%%%%%%%%%%%%%%%%%%%%%%%%%%%%%%%%%%%%%%%%%%%%%%%%%%%%%
%
%  Acknowledgement
%
%%%%%%%%%%%%%%%%%%%%%%%%%%%%%%%%%%%%%%%%%%%%%%%%%%%%%%%%%%%%%%%%%%%%%%%%%%%%%%%%%%%%%%%%%%%%%%%%%%%%%%%%%%%%%%%%%%%%%%

S.-A.\ B. acknowledges support from Heisenberg Programme of the Deutsche Forschungsgemeinschaft (DFG, German Research Foundation) under the project No. 404073166. 

\appendix

\bibliographystyle{unsrt}
%\bibliography{literaturhoney}

%\begin{thebibliography}{999}  
%\end{thebibliography}
\end{document}